# Quantum States Estimation: Root Approach

Yu. I. Bogdanov

OAO "Angstrem", 124460, Moscow, Russia[1]

**ABSTRACT**

Multiparametric statistical model providing stable reconstruction of parameters by observations is considered. The only general method of this kind is the root model based on the representation of the probability density as a squared absolute value of a certain function, which is referred to as a psi-function in analogy with quantum mechanics. The psi-function is represented by an expansion in terms of an orthonormal set of functions. It is shown that the introduction of the psi-function allows one to represent the Fisher information matrix as well as statistical properties of the estimator of the state vector (state estimator) in simple analytical forms. A new statistical characteristic, a confidence cone, is introduced instead of a standard confidence interval. The chi-square test is considered to test the hypotheses that the estimated vector converges to the state vector of a general population and that both samples are homogeneous. The expansion coefficients are estimated by the maximum likelihood method. The method proposed may be applied to its full extent to solve the statistical inverse problem of quantum mechanics (root estimator of quantum states). In order to provide statistical completeness of the analysis, it is necessary to perform measurements in mutually complementing experiments (according to the Bohr terminology). The maximum likelihood technique and likelihood equation are generalized in order to analyze quantum mechanical experiments. It is shown that the requirement for the expansion to be of a root kind can be considered as a quantization condition making it possible to choose systems described by quantum mechanics from all statistical models consistent, on average, with the laws of classical mechanics.

## 1. FISHER INFORMATION MATRIX AND STATE ESTIMATOR

For the sake of simplicity, consider first a real valued psi function. Let an expansion have the form

$$\psi(x) = \sqrt{1 - (c_1^2 + \ldots + c_{s-1}^2)}\varphi_0(x) + c_1\varphi_1(x) + \ldots + c_{s-1}\varphi_{s-1}(x). \quad (1)$$

Here, we have eliminated the coefficient $c_0 = \sqrt{1 - (c_1^2 + \ldots + c_{s-1}^2)}$ from the set of parameters to be estimated, since it is expressed via the other coefficients by the normalization condition.

The parameters $c_1, c_2, \ldots, c_{s-1}$ are independent. We will study their asymptotic behavior using the Fisher information matrix [1-2]

$$I_{ij}(c) = n \cdot \int \frac{\partial \ln p(x,c)}{\partial c_i} \frac{\partial \ln p(x,c)}{\partial c_j} p(x,c) dx$$

The fundamental significance of the Fisher information matrix consists in its property to set the constraint on achievable (in principle) accuracy of statistical estimators. According to the Cramer - Rao inequality [1-2], the matrix $\Sigma(\hat{\theta}) - I^{-1}(\theta)$ is nonnegative for any unbiased estimator $\hat{\theta}$ of an unknown vector valued parameter $\theta$. Here, $\Sigma(\hat{\theta})$ is the covariance matrix for the estimator $\hat{\theta}$. The corresponding difference asymptotically tends to a zero matrix for the maximum likelihood estimators (asymptotic efficiency).

It is of particular importance for our study that the Fisher information matrix drastically simplifies if the psi function is introduced [3-4]

---

[1] e-mail: bogdanov@angstrem.ru

$$I_{ij} = 4n \cdot \int \frac{\partial \psi(x,c)}{\partial c_i} \frac{\partial \psi(x,c)}{\partial c_j} dx. \qquad (2)$$

In the case of the expansion (1), the information matrix $I_{ij}$ is $(s-1) \times (s-1)$ matrix of the form

$$I_{ij} = 4n\left(\delta_{ij} + \frac{c_i c_j}{c_0^2}\right), \quad c_0 = \sqrt{1 - (c_1^2 + ... + c_{s-1}^2)}. \qquad (3)$$

A noticeable feature of the expression (3) is its independence on the choice of basis functions. Note that only the representation of the density in the form $p = |\psi|^2$ results in a universal (and simplest) structure of the Fisher information matrix.

In view of the asymptotic efficiency of the maximum likelihood estimators, the covariance matrix of the state estimator is the inverse Fisher information matrix:

$$\Sigma(\hat{c}) = I^{-1}(c) \qquad (4)$$

Let us extend the covariance matrix by appending the covariance between the $c_0$ component of the state vector and the other components. In result, we find that the covariance matrix components are

$$\Sigma_{ij} = \frac{1}{4n}(\delta_{ij} - c_i c_j) \qquad i,j = 0,1,...,s-1. \qquad (5)$$

From the geometrical standpoint, the covariance matrix (5) is a second-order tensor. Moreover, the covariance matrix (up to a constant factor) is a single second-order tensor satisfying the normalization condition.

In quantum mechanics, the matrix

$$\rho_{ij} = c_i c_j \qquad (6)$$

is referred to as a density matrix (of a pure state). Thus,

$$\Sigma = \frac{1}{4n}(E - \rho), \qquad (7)$$

where $E$ is the $s \times s$ unit matrix.

In the diagonal representation,

$$\Sigma = UDU^+, \qquad (8)$$

where $U$ and $D$ are unitary (orthogonal) and diagonal matrices, respectively.

As is well known from quantum mechanics, the density matrix of a pure state has the only (equal to unity) element in the diagonal representation. Thus, in our case, the diagonal of the $D$ matrix has the only element equal to zero (the corresponding eigenvector is the state vector); whereas the other diagonal elements are equal to $\frac{1}{4n}$ (corresponding eigenvectors and their linear combinations form a subspace that is orthogonal complement to the state vector). The zero element at a principle diagonal indicates that the inverse matrix (namely, the Fisher information matrix of the $s$-th order) does not exist. It is clear since there are only $s-1$ independent parameters in the distribution.

The results on statistical properties of the state vector reconstructed by the maximum likelihood method can be summarized as follows. In contrast to a true state vector, the estimated one involves noise in the form of a random deviation vector located in the space orthogonal to the true state vector. The components of the deviation vector (totally,

$s-1$ components) are asymptotically normal independent random variables with the same variance $\frac{1}{4n}$. In the aforementioned $s-1$-dimensional space, the deviation vector has an isotropic distribution, and its squared length is the random variable $\frac{\chi^2_{s-1}}{4n}$, where $\chi^2_{s-1}$ is the random variable with the chi-square distribution of $s-1$ degrees of freedom, i.e.

$$1-\left(c,c^{(0)}\right)^2 = \frac{\chi^2_{s-1}}{4n}. \tag{9}$$

This expression means that the squared scalar product of the true and estimated state vectors is smaller than unity by asymptotically small random variable $\frac{\chi^2_{s-1}}{4n}$.

The results found allow one to introduce a new stochastic characteristic, namely, a confidence cone (instead of a standard confidence interval). Let $\vartheta$ be the angle between an unknown true state vector $c^{(0)}$ and that $c$ found by solving the likelihood equation. Then,

$$\sin^2 \vartheta = 1 - \cos^2 \vartheta = 1 - \left(c,c^{(0)}\right)^2 = \frac{\chi^2_{s-1}}{4n} \leq \frac{\chi^2_{s-1,\alpha}}{4n}. \tag{10}$$

Here, $\chi^2_{s-1,\alpha}$ is the quantile corresponding to the significance level $\alpha$ for the chi-square distribution of $s-1$ degrees of freedom.

The set of directions determined by the inequality (10) constitutes the confidence cone. The axis of a confidence cone is the reconstructed state vector $c$. The confidence cone covers the direction of an unknown state vector at a given confidence level $P=1-\alpha$.

## 2. STATISTICAL ANALYSIS OF MUTUALLY COMPLEMENTING EXPERIMENTS

We have defined the psi function as a complex-valued function with the squared absolute value equal to the probability density. From this point of view, any psi function can be determined up to arbitrary phase factor $\exp(iS(x))$. In particular, the psi function can be chosen real-valued.

At the same time, from the physical standpoint, the phase of psi function is not redundant. The psi function becomes essentially complex valued function in analysis of mutually complementing (according to Bohr) experiments with micro objects [5].

According to quantum mechanics, experimental study of statistical ensemble in coordinate space is incomplete and has to be completed by study of the same ensemble in another (canonically conjugate, namely, momentum) space. Note that measurements of ensemble parameters in canonically conjugate spaces (e.g., coordinate and momentum spaces) cannot be realized in the same experimental setup.

The uncertainty relation implies that the two-dimensional density in phase space $P(x,p)$ is physically senseless, since the coordinates and momenta of micro objects cannot be measured simultaneously. The coordinate $P(x)$ and momentum $\widetilde{P}(p)$ distributions should be studied separately in mutually complementing experiments and then combined by introducing the psi function.

The coordinate-space and momentum-space psi functions are related to each other by the Fourier transform

$$\psi(x) = \frac{1}{\sqrt{2\pi}} \int \widetilde{\psi}(p) \exp(ipx) dp, \quad \widetilde{\psi}(p) = \frac{1}{\sqrt{2\pi}} \int \psi(x) \exp(-ipx) dx. \tag{11}$$

Consider a problem of estimating an unknown psi function ($\psi(x)$ or $\widetilde{\psi}(p)$) by experimental data observed both in coordinate and momentum spaces. We will refer to this problem as an statistical inverse problem of quantum mechanics [6-9] (do not confuse it with an inverse problem in the scattering theory). The predictions of quantum mechanics are considered as a direct problem. Thus, we consider quantum mechanics as a stochastic theory, i.e., a theory describing statistical (frequency) properties of experiments with random events. However, quantum mechanics is a special stochastic theory, since one has to perform mutually complementing experiments (space-time description has to be completed by momentum-energy one) to get statistically full description of a population (ensemble). In order for various representations to be mutually consistent, the theory should be expressed in terms of probability amplitude rather than probabilities themselves.

Methodologically, the method considered here essentially differs from other well known methods for estimating quantum states that arise from applying the methods of classical tomography and classical statistics to quantum problems [10-12]. The quantum analogue of the distribution density is the density matrix and the corresponding Wigner distribution function. Therefore, the methods developed so far have been aimed at reconstructing the aforementioned objects in analogy with the methods of classical tomography (this resulted in the term "quantum tomography") [13].

In [14], a quantum tomography technique on the basis of the Radon transformation of the Wigner function was proposed. The estimation of quantum states by the method of least squares was considered in [15]. The maximum likelihood technique was first presented in [16,17]. The version of the maximum likelihood method providing fulfillment of basic conditions imposed of the density matrix (hermicity, nonnegative definiteness, and trace of matrix equal to unity) was given in [18,19]. Characteristic features of all these methods are rapidly increasing calculation complexity with increasing number of parameters to be estimated and ill-posedness of the corresponding algorithms, not allowing one to find correct stable solutions.

The orientation toward reconstructing the density matrix overshadows the problem of estimating more fundamental object of quantum theory, i.e., the state vector (psi function). Formally, the states described by the psi function are particular cases of those described by the density matrix. On the other hand, this is the very special case that corresponds to fundamental laws in Nature and is related to the situation when the state described by a large number of unknown parameters may be stable and estimated up to the maximum possible accuracy.

Let us consider generalization of the maximum likelihood principle and likelihood equation for estimation of the state vector of a statistical ensemble on the basis of experimental data obtained in mutually complementing experiments. To be specific, we will assume that corresponding experiments relate to coordinate and momentum spaces.

We define the likelihood function as

$$L(x,p|c) = \prod_{i=1}^{n} P(x_i|c) \prod_{j=1}^{m} \widetilde{P}(p_j|c). \qquad (12)$$

Here, $P(x_i|c)$ and $\widetilde{P}(p_j|c)$ are the densities in mutually complementing experiments corresponding to the same state vector $c$. We assume that $n$ measurements were made in the coordinate space; and $m$, in the momentum one.

Then, the log likelihood function has the form

$$\ln L = \sum_{i=1}^{n} \ln P(x_i|c) + \sum_{j=1}^{m} \ln \widetilde{P}(p_j|c). \qquad (13)$$

The maximum likelihood principle together with the normalization condition evidently results in the problem of maximization of the following functional:

$$S = \ln L - \lambda(c_i c_i^* - 1), \qquad (14)$$

where $\lambda$ is the Lagrange multiplier and

$$\ln L = \sum_{k=1}^{n} \ln(c_i c_j^* \varphi_i(x_k) \varphi_j^*(x_k)) + \sum_{l=1}^{m} \ln(c_i c_j^* \widetilde{\varphi}_i(p_l) \widetilde{\varphi}_j^*(p_l)). \qquad (15)$$

Here, $\widetilde{\varphi}_i(p)$ is the Fourier transform of the function $\varphi_i(x)$.

Hereafter, we imply the summation over recurring indices numbering the terms of the expansion in terms of basis functions. On the contrary, statistical sums denoting the summation over the sample points will be written in an explicit form.

The necessary condition $\frac{\partial S}{\partial c_i^*} = 0$ for an extremum yields the likelihood equation

$$R_{ij} c_j = \lambda c_i \quad i,j = 0,1,\ldots,s-1, \qquad (16)$$

where the $R$ matrix is determined by

$$R_{ij} = \sum_{k=1}^{n} \frac{\varphi_i^*(x_k)\varphi_j(x_k)}{P(x_k)} + \sum_{l=1}^{m} \frac{\widetilde{\varphi}_i^*(p_l)\widetilde{\varphi}_j(p_l)}{\widetilde{P}(p_l)}. \qquad (17)$$

The problem (16) is formally linear. However, the matrix $R_{ij}$ depends on an unknown densities $P(x)$ and $\widetilde{P}(p)$. Therefore, the problem under consideration is actually nonlinear, and should be solved by the iteration method [6-8]. An exception is the histogram density estimator when the problem can be solved straightforwardly.

Multiplying both parts of Eq. (16) by $c_i^*$ and summing with respect to $i$, we find that the most likely state vector $c$ always corresponds to its eigenvalue $\lambda = n + m$ of the $R$ matrix (equal to sum of measurements).

An optimal number of harmonics in the expansion is appropriate to choose, on the basis of the compromise, between two opposite tendencies: the accuracy of the estimation of the function approximated by a finite series increases with increasing number of harmonics, however, the statistical noise level also increases.

The likelihood equation in the root state estimator method has a simple quasilinear structure and admits developing an effective fast-converging iteration procedure even in the case of multiparametric problems. The numerical implementation of the proposed algorithm is considered by the use of the set of Chebyshev-Hermite functions as a basis set of functions [6-7].

The implication of the root estimator method to statistical reconstruction of optical quantum states is considered in [9].

Examples of mutually complementing experiments that are of importance from the physical point of view are diffraction patterns (for electrons, photons, and any other particles) in the near-field zone (directly downstream of the diffraction aperture) and in the Fraunhofer zone (far from the diffraction aperture). The intensity distribution in the near-field zone corresponds to the coordinate probability distribution; and that in the Fraunhofer zone, the momentum distribution. The psi function estimated by these two distributions describes the wave field (amplitude and phase) directly at the diffraction aperture. The psi function dynamics described by the Schrödinger equation for particles and the Leontovich parabolic equation for light allows one to reconstruct the whole diffraction pattern (in particular, the Fresnel diffraction).

In the case of a particle subject to a given potential (e.g., an atomic electron) and moving in a finite region, the coordinate distribution is the distribution of the electron cloud, and the momentum distribution is detected in a thought experiment where the action of the potential abruptly stops and particles move freely to infinity.

In quantum computing, the measurement of the state of a quantum register corresponds to the measurement in coordinate space; and the measurement of the register state after performing the discrete Fourier transform, the measurement in momentum space. A quantum register involving $n$ qubits can be in $2^n$ states; and correspondingly, the same number of complex parameters is to be estimated. Thus, exponentially large number of measurements of identical registers is required to reconstruct the psi function if prior information about this function is lacking.

The state of quantum register is determined by the psi function

$$\psi = c_i |i\rangle \tag{18}$$

The probability amplitudes in the conjugate space corresponding to complementing measurements are

$$\widetilde{c}_i = U_{ij} c_j \tag{19}$$

The likelihood function relating to $n + m$ mutually complementing measurements is

$$L = \prod_i \left(c_i c_i^*\right)^{n_i} \prod_j \left(\widetilde{c}_j \widetilde{c}_j^*\right)^{m_j} \tag{20}$$

Here, $n_i$ and $m_j$ are the number of measurements made in corresponding states.

In the case under consideration, the likelihood equation similar to (16) has the form

$$\frac{1}{n+m}\left[\frac{n_i}{c_i^*} + \sum_j \frac{m_j U_{ji}^*}{\widetilde{c}_j^*}\right] = c_i \tag{21}$$

Figure 1 shows the comparison between exact densities that could be calculated if the psi function of an ensemble is known (solid line), and estimators obtained in mutually complementing experiments (quantum register: 8 qubits, $2^8 = 256$ states). In each experiment, the sample size is 10000 points. In Fig. 2, the exact psi function is compared to that estimated by samples.

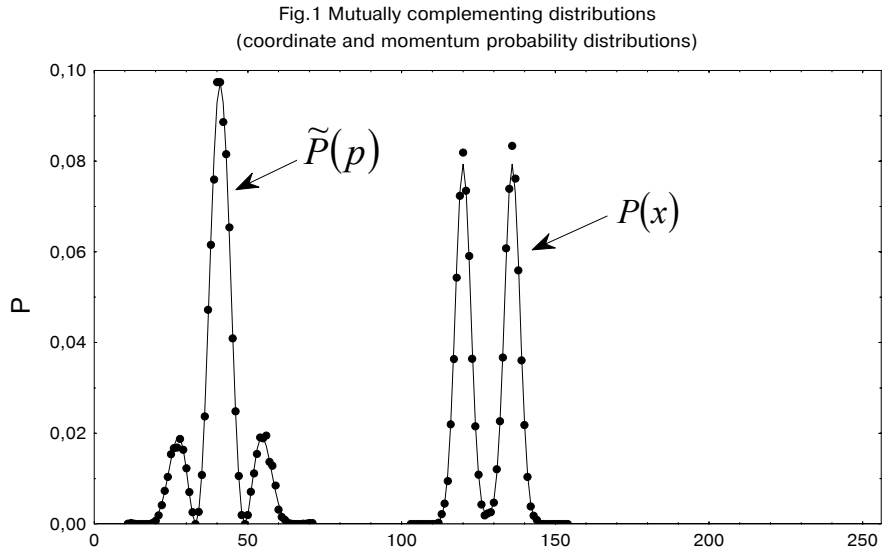

Fig.1 Mutually complementing distributions
(coordinate and momentum probability distributions)

## 3. ROOT ESTIMATOR AND QUANTUM DYNAMICS

Assume that the mechanical equations are satisfied only for statistically averaged quantities (the averaged Newton's second law of motion)

$$\frac{d^2}{dt^2}\left(\int P(x)\vec{x}\, dx\right) = -\frac{1}{m}\left(\int P(x)\frac{\partial U}{\partial \vec{x}}\, dx\right) \tag{22}$$

Let us require the density $P(x)$ to admit the root expansion [8], i.e.,

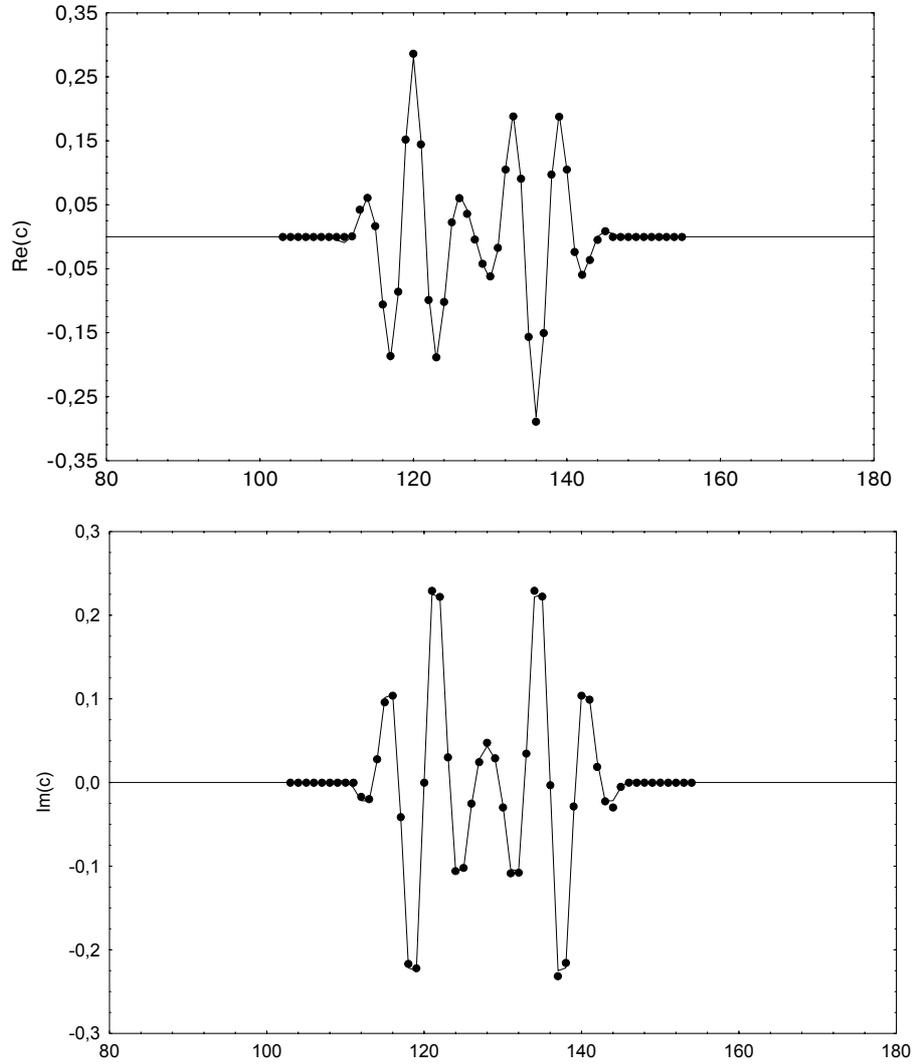

Fig. 2 Comparison between exact psi- function (solid line) and that estimated by a sample (dots)

$$P(x) = |\psi(x)|^2, \qquad (23)$$

where

$$\psi(x) = c_j(t)\varphi_j(x) \qquad (24)$$

We will search for the time dependence of the expansion coefficients in the form of harmonic dependence

$$c_j(t) = c_{j0} \exp(-i\omega_j t). \qquad (25)$$

Then, Eq. (22) yields

$$m(\omega_j - \omega_k)^2 c_{j0} c_{k0}^* \langle k|\vec{x}|j\rangle \exp(-i(\omega_j - \omega_k)t) =$$
$$= c_{j0} c_{k0}^* \langle k|\frac{\partial U}{\partial \vec{x}}|j\rangle \exp(-i(\omega_j - \omega_k)t) \tag{26}$$

Here, the summation over recurring indices $j$ and $k$ is meant. The matrix elements in (26) are determined by the formulas

$$\langle k|\vec{x}|j\rangle = \int \varphi_k^*(x) \vec{x}\, \varphi_j(x) dx \tag{27}$$

$$\langle k|\frac{\partial U}{\partial \vec{x}}|j\rangle = \int \varphi_k^*(x) \frac{\partial U}{\partial \vec{x}} \varphi_j(x) dx \tag{28}$$

In order for the expression (26) to be satisfied at any instant of time for arbitrary initial amplitudes, the left and right sides are necessary to be equal for each matrix element. Therefore,

$$m(\omega_j - \omega_k)^2 \langle k|\vec{x}|j\rangle = \langle k|\frac{\partial U}{\partial \vec{x}}|j\rangle \tag{29}$$

This expression is a matrix equation of the Heisenberg quantum dynamics in the energy representation (written in the form similar to that of the Newton's second law of motion). The basis functions and frequencies satisfying (29) are the stationary states and frequencies of a quantum system, respectively (in accordance with the equivalence of the Heisenberg and Schrödinger pictures).

Indeed, let us construct the diagonal matrix from the system frequencies $\omega_j$. The matrix under consideration is Hermitian, since the frequencies are real numbers. This matrix is the representation of a Hermitian operator with eigenvalues $\omega_j$, i.e.,

$$H|j\rangle = \hbar \omega_j |j\rangle \tag{30}$$

Let us find an explicit form of this operator. In view of (30), the matrix relationship (29) can be represented in the form of the operator equation

$$[H[Hx]] = \frac{\hbar^2}{m} \hat{\partial} U, \tag{31}$$

where $\hat{\partial} = \frac{\partial}{\partial x}$ is the operator of differentiation and $[\ ]$, the commutator.

The Hamiltonian of a system

$$H = -\frac{\hbar^2}{2m} \hat{\partial}^2 + U(x) \tag{32}$$

is the solution of operator equation (31).

Thus, if the root density estimator is required to satisfy the averaged classical equations of motion, the basis functions and frequencies of the root expansion cannot be arbitrary, but have to be eigenfunctions and eigenvalues of the system Hamiltonian, respectively.

The relationships providing that the averaged equations of classical mechanics are satisfied for quantum systems are referred to as the Ehrenfest equations [20]. These equations are insufficient to describe quantum dynamics. As it has been shown above, an additional condition allowing one to transform a classical system into the quantum one (i.e., quantization condition) is actually the requirement for the density to be of the root form.

Thus, if we wish to turn from the rigidly deterministic (Newtonian) description of a dynamical system to the statistical one, it is natural to use the root expansion of the density distribution to be found, since only in this case a stable statistical model can be found. On the other hand, the choice of the root expansion basis determined by the eigenfunctions of the energy operator (Hamiltonian) is not simply natural, but the only possible way consistent with the dynamical laws.

## 4. CONCLUSIONS

Let us state a short summary.

Search for multiparametric statistical model providing stable estimation of parameters on the basis of observed data results in constructing the root density estimator. The root density estimator is based on the representation of the probability density as a squared absolute value of a certain function, which is referred to as a psi-function in analogy with quantum mechanics. The method proposed is an efficient tool to solve the basic problem of statistical data analysis, i.e., estimation of distribution density on the basis of experimental data.

The coefficients of the psi-function expansion in terms of orthonormal set of functions are estimated by the maximum likelihood method providing optimal asymptotic properties of the method (asymptotic unbiasedness, consistency, and asymptotic efficiency). The likelihood equation in the root density estimator method has a simple quasilinear structure and admits developing an effective fast-converging iteration procedure even in the case of multiparametric problems.

The introduction of the psi-function allows one to represent the Fisher information matrix as well as statistical properties of the sate vector estimator in simple analytical forms. Basic objects of the theory (state vectors, information and covariance matrices etc.) become simple geometrical objects in the Hilbert space that are invariant with respect to unitary (orthogonal) transformations.

A new statistical characteristic, a confidence cone, is introduced instead of a standard confidence interval. The chi-square test is considered to test the hypotheses that the estimated vector equals to the state vector of general population and that both samples are homogeneous.

The root state estimator may be applied to analyze the results of experiments with micro objects as a natural instrument to solve the inverse problem of quantum mechanics: estimation of psi function by the results of mutually complementing (according to Bohr) experiments. Generalization of the maximum likelihood principle to the case of statistical analysis of mutually complementing experiments is proposed.

It is shown that the requirement for the density to be of the root form is the quantization condition. Actually, one may say about the root principle in statistical description of dynamic systems. According to this principle, one has to perform the root expansion of the distribution density in order to provide the stability of statistical description. On the other hand, the root expansion is consistent with the averaged laws of classical mechanics when the eigenfunctions of the energy operator (Hamiltonian) are used as basis functions. Figuratively speaking, there is no a regular statistical method besides the root one, and there is no regular statistical mechanics besides the quantum one.